\documentclass[12pt]{article}
\usepackage{rotating}
\usepackage{epsfig,amssymb}
\input{epsf}

\def\laq{\raise 0.4 ex \hbox{$<$}\kern -0.8 em\lower 0.62 ex\hbox{$\sim$}}
\def\gaq{\raise 0.4 ex \hbox{$>$}\kern -0.7 em\lower 0.62 ex\hbox{$\sim$}}

\def\beq{\begin{equation}}
\def\eeq{\end{equation}}
\def\beqa{\begin{eqnarray}}
\def\eeqa{\end{eqnarray}}

\def\PL{{\it Phys. Lett.} }

\def\PR{{\it Phys. Rev.} }

 \def\frac#1#2{{\textstyle{{#1}\over {#2}}}}
\def\ni{\noindent} \def\lsim{\mathrel{\rlap{\lower4pt\hbox{\hskip1pt$\sim$}}
    \raise1pt\hbox{$<$}}} \def\gsim{\mathrel{\rlap{\lower4pt\hbox{\hskip1pt$\sim$}}
    \raise1pt\hbox{$>$}}}
\def\sqr#1#2{{\vcenter{\vbox{\hrule height.#2pt
         \hbox{\vrule width.#2pt height#1pt \kern#1pt
         \vrule width.#2pt}
         \hrule height.#2pt}}}}

\def\gappeq{\mathrel{\rlap {\raise.5ex\hbox{$>$}} {\lower.5ex\hbox{$\sim$}}}}

\def\lappeq{\mathrel{\rlap{\raise.5ex\hbox{$<$}}
{\lower.5ex\hbox{$\sim$}}}}

\begin{document}
\pagestyle{plain}

\begin{flushright}
January 2010
\end{flushright}
\vspace{15mm}

\begin{center}

{\Large\bf Cosmological thinking: cultural heritage and challenge$^*$}

\vspace*{1.0cm}

Orfeu Bertolami$^{**}$ \\
\vspace*{0.5cm}
{Instituto Superior T\'ecnico, Departamento de F\'\i sica, \\
Av. Rovisco Pais, 1049-001 Lisboa, Portugal}\\

\vspace*{2.0cm}
\end{center}

\begin{abstract}

\noindent
The limitations of current technology do not allow one to foresee the expansion of the humankind beyond our planet for at least a 
few decades. Furthermore, the
laws of physics, as for as they are known, preclude any form of traveling beyond the speed of light, as well as any 
viable and stable space-time shortcuts 
(wormholes, warp-drives, etc) that would facilitate cosmic traveling. Given the vastness of the Universe these are insurmountable 
obstacles for any {\it in situ} exploration 
of the cosmos beyond our most immediate cosmic neighbourhood. Nevertheless, the Universe is transparent and contains countless 
sources of visible light. Actually, in the last decades, technological 
developments have made possible to observe the cosmos throughout most of the electromagnetic spectrum as well as to perform 
dynamical studies that allow 
perceiving the presence of invisible components such as black holes, dark matter and dark energy. In this respect, humankind 
has then been given the opportunity 
to unravel the inner workings of the cosmos and through this process be part of the cosmic habitat. In this contribution 
various forms of cosmological thinking will 
be discussed, from some myths of creation till some of the latest scientific discoveries.

\end{abstract}

\vfill
\noindent\underline{\hskip 140pt}\\[4pt]
{$^*$ Talk delivered at the Third International Symposium "Fronteiras da Ci\^encia: A Humanidade e o Cosmos", 
13 - 14 November 2009, Universidade Fernando Pesssoa, Oporto, Portugal.} \\
\noindent
{$^{**}$ Also at Instituto de Plasmas e Fus\~ao Nuclear, Instituto Superior T\'ecnico,
Lisboa, Portugal.} \\
{E-mail address: orfeu@cosmos.ist.utl.pt}

\newpage
\section{Cosmological Thinking: From the Myths of Creation to the Big Bang}
\label{sec:intro}
 
It is somewhat unusual that a theoretical physicist discusses issues that are related with history and the history of 
ideas given that these matters go beyond 
his field of expertise. However, this is somewhat inevitable when one's purpose is to discuss the underlying assumptions and 
the challenges that a physicist faces in analyzing the impact on human culture 
of most recent discoveries in  cosmology, and physics in general. Furthermore, given the nature of such an interesting meeting 
and the fact that one of the main motivations of the International 
Year of Astronomy is to discuss the role that astronomy and cosmology has on our world, this author cannot refrain from leaving 
the ground of his expertise in order to discuss the role that the cosmological reasoning has played and 
will continue to play on the cultural development of humankind.  

Most likely, astronomy is the most ancient of all sciences thanks to the fact that the Universe shines magnificently in the 
optical region of the electromagnetic spectrum and that our atmosphere 
is transparent in this range wavelengths. This allowed our ancestors to observe the cosmos and wonder about its beauty 
and inner workings since early on. Indeed, the first astronomical instruments, stone circles and stone constructions, 
used to guide human activities with the apparent motion of the sun, date back to about 2000-2500 B.C. 

It is useful, in order to put things under perspective, to point out that civilization has emerged after 
the Neolithic Revolution in the Near East, about 15 thousand years ago, and the ensued appearance of 
city states about 10 thousand years ago. Civilization, as we know it, involves the emergence of a 
form of kingdom and the development of a writing system. Evidence shows that  
the latter took place about 6 thousand years ago \cite{Cook}.       

Speculative thinking about the 
nature, the dimension, the duration and of our standing in the cosmos was a central concern of the first civilizations 
in Mesopotamia, Egypt, India, China and Crete.  
Later, at about 400 -- 350 B.C., Hellenic thinking, most particularly through the systematical approach of Aristotle (384 -- 322 B.C.) to 
all matters of the physical and philosophical world, put Earth at the center of the Universe. Consensus about this view was challenged 
by Aristarchus (310 -- ca. 230 B.C.) of Samos, the Greek astronomer and mathematician, who suggested that the Sun should be 
at the center of the solar system 1800 years earlier than Copernicus (1473 -- 1543). 

But before diving into the modern view of the cosmos and its impact of human thinking, it is worth discussing 
the very origin of the 
cosmological thinking. Most of the ideas discussed in this section can be found in Ref. \cite{Bertolami06a} 
and in the bibliography therein. To start with, one could not fail to note  
that any civilization, when reflecting upon its standing in the world traces back its historical roots, 
and in the process it speculates on its origin, creating for that a cosmogony. 
This historical reconstruction about the origin is usually encapsulated into a myth, a myth of creation. These myths 
treat the Universe as a living being and connect the creation process to the divine intervention, emphasizing in this way the 
supernatural and the extraordinary. Myths 
of creation range from the most original, when pristine, to somewhat rich in influences of the 
neighbouring civilizations, often more powerful and ancient. 
Anthropocentric reasoning is a fundamental principle in most of the myths of creation.
Myths of creation of the world are conveyed in an epic and solemn language 
and most often start with the transformation of order out the chaos and the 
separation of the sky and the earth. In opposition, the creation of humankind features more mundane deeds 
and often involves some sort of copulation or masturbation.

Another common feature among the myths of creation is the concept of design with a purpose, which is 
encapsulated by the {\it Word}, the divine code or recipe for creation. This are encountered, for instance, in the 
Sumerian cosmogony, in the Genesis of the Hebrew Torah, in the geometrical universe of the Greeks, etc. These features 
can also be found in completely different cultural contexts, as exemplified by the myth of creation of the Guarani Indians, 
who lived throughout Brazil and South America till recent centuries. The god Namandu, the true father, 
the one who knows the meaning of the creative Word\footnote{Translation due to the author.}:

\ni
``...

In his divine knowledge of things,

\ni   
knowledge that unfolds things,

\ni
he knows by himself

\ni
the source of what is supposed to be gathered;

\ni
Earth is not yet,

\ni 
the original night remains,

\ni
the knowledge of things is not known yet,

\ni 
but he by himself

\ni
the source whose fate is to be gathered."

Actually, a myth must be regarded as the most basic form of savage (pristine) thinking, 
and despite the cultural differences, the common features of the myths, and of the myths of creation in 
particular, can be understood according to the anthropologist Claude L\'evi-Strauss 
\cite{LeviSrauss}, by the fact that the myths have an essential kern of irreducible and unchanging elements, 
the {\it mythemes}. {\it Mytheme} are the fundamental units of myth. These minimal units are reassembled in 
various ways and linked in more complicated relationships, likewise 
a molecule in a compound. The relation between mytheme is that of binary (dialectic) opposition.  

In order to close the discussion on the myths of creation, one brief describes a 
form of mythical thinking that is still active, and that connects a whole range of cultural and ritualistic 
activities, the {\it Dreaming} of the Australian Aborigines.  

For the Australian Aborigines, 
the ancestral beings moved across the land creating life and the important geographical features of the landscape. 
This creative power is revealed through the {\it Dreaming}. {\it Dreaming}, sometimes also referred to as {\it Dreamtime}, 
as it is translated from the Arrernte language, also means ``to see and to understand the big law". {\it Dreaming} stories pass 
on the knowledge, the cultural values and the belief system of the Australian Indigenous people to later generations. 
It should be pointed out that Australian Aborigines have most likely the longest continuous cultural history of all humans groups. 
The history of this human group is estimated to date back between 50,000 and 65,000 years. Before the European settlement 
of Australia, there were around 600 different Aboriginal nations, based on language groups. Their rich cultural heritage connects the 
ancestral subjective experience with the objective world.

{\it Dreaming} stories manifest themselves through song, dance, painting \cite{Meyers, Benjamin} 
and storytelling and allow the Aborigines to maintain a bond 
with their cultural lineage from ancient times till present. 
Even though, most often {\it Dreamings} refers to somewhat mundane occurrences and manifestations of nature, seduction, wandering, 
flood, thundering and so on, it is rather remarkable that they might also involve 
a ``cosmic" connection. A quite representative example is the {\it Milky Way Dreaming} which materializes in a 
pictorial form through the strikingly beautiful {\it Seven Sisters} 
canvas of the Australian artist Grabiela Possum Nungarrayi (http://www.authaboriginalart.com.au/index.asp), a painting 
which gave origin to a series of painting representing the Pleaides constellation (Figure 1) from 1998 onwards.  

The background story of this Dreamtime involves the chasing of a group of seven Napaltjarri spirit 
women by a Tjakamara spirit man who had noticed them occupied in ceremonial singing and dancing. In the words of Gabriella 
Possum Nungarrayi: 

\ni
``The sisters could feel nothing else but their own music and dancing as if they were hypnotized. Like a corroboree. 
Then they saw a bright light in the distance and knew it was the man coming."
The Tjakamara man hid nearby and tried to use ``love magic" to attract them. The magic did not work on the young 
women and none of them had any intention of spending time with the man. They ran away into the sky to cluster into 
seven stars known as the Pleiades in the constellation of Taurus. The Tjakamara man followed them and became the 
Morning Star in Orion's belt. As the constellations move across the night sky in front of the 
Milky Way the sisters can be seen flying ahead of the Tjakamara man who can never catch up with them. The sisters 
come back to earth through Dreaming ceremonies performed by Aboriginal women and then return 
to their eternal freedom in the sky where they keep watch over their earthly sisters. 

The artist also states: ``People think we don't have writing. Our writing is on the ground and all around us. 
The old people can see the signs when the sisters have been back.''
This is all quite remarkable as it shows that the Australian Aborigines 
still keep a close bond with their ancestral culture and hence with their land and with the cosmos itself. 

\begin{figure}[]
\centering \includegraphics[width=0.5\textwidth]{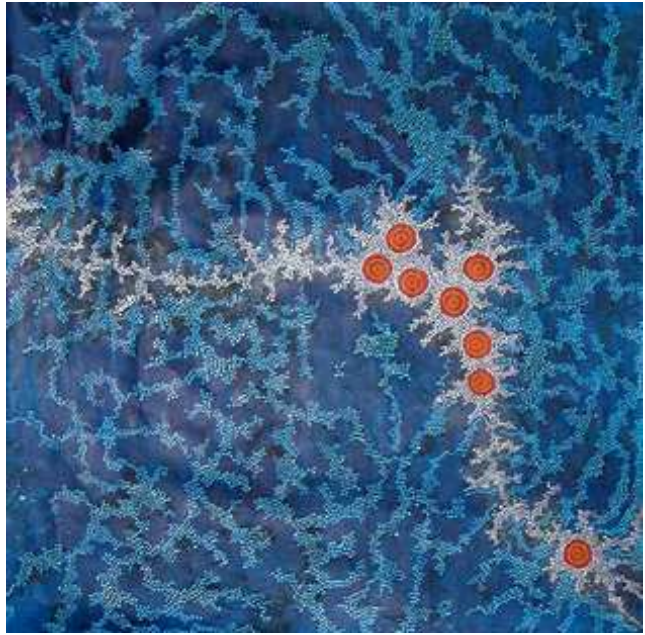}

\caption{``Milk Way Dreaming''  by Grabiela Possum Nungarrayi}
\label{fig:dark} 
\end{figure}

\ni 
Carrying our discussion further on, it should be remembered that the emergence of 
the so-called rational religions, {\it circa} 3000 B.C., associated the creation 
to the divine and actually the regularity of the natural phenomena got attributed to the 
judgment, mood, choices and wishes of the gods. Hesiod (VII-VI B.C.), the Greek poet, was the 
first to rationalize myth and nature through the process of semantic duplication: the name of the 
gods were used to denote the natural phenomena they were the incarnation of. According to Hesiod, 
the origin of the world corresponded to the birth of the gods.

In ancient times, physics was considered in the context of philosophy, it was the natural philosophy. 
The cosmos encompassed the world, the city-states and their citizens. In its development, physics broke free from philosophy 
and that brought the origin of the world to the realm of natural phenomena, which can be studied and 
subjected to the same rationalization and methods used to analyse nature. 

Although cosmic thinking was an important part of rationalization of the world and of the sky mechanics in the Hellenic 
period, it was only after the Copernican revolution, and more particularly with the scientific revolution in the 
XVII century, that cosmological speculation could be completely freed from the theological and philosophical burden it was
subjected so far. Galileo observations, Newton's mechanics and gravitation are 
universal and describe to the whole cosmos. Its universal prevalence 
allows one to speculate about the ``theory of the sky'' as developed 
by Kant (``Algemeine Naturgeschichte und Theorie des Himmels'', 1755) and Laplace (``Sist\`eme du monde'', 1810). 
This scientific revolution initiated of a 
cultural tradition one still lives on. For Kant, Laplace and followers, Newton's mechanics could explain  
the origin of the solar system and of all planetary systems. The laws of the ``universal clock'' would keep, through the 
conservation of energy, linear and angular momenta, the Universe working without the need of 
any supernatural intervention. This conceptual expansion was closely followed 
by an expansion of the astronomical observation. Indeed, the observations by Herschel in late 
the XVIII century and the early XIX century, suggested that ``nebulae'' were quite generic and widespread, and that the Milk 
Way was a complex and quite rich system of stars, gas and dusk (and dark matter as it is currently known),  but, in any case, 
a single element in a vast Universe composed by a great number of similar systems. 
That these systems, galaxies, are independent from each other was shown by Hubble in 1923.  

Thus, scientific developments shifted definitely the discussion about the 
origin of the world towards science. The following step was given by Einstein 
when he proposed in 1917 the first cosmological model based on general relativity, a model where the universe was 
a static and eternal. Thus, theoretical cosmology has emerged from the understanding that the general theory of relativity was the 
suitable framework to address the cosmological problem, given the that this theory concerns the global nature of the 
four dimensional space-time (see Ref. \cite{Bertolami08a} for a discussion of this historical development). 
In the context of the general theory of relativity, space-time acquires a plasticity that is shaped by matter configurations and 
can therefore behave in quite unusual and unfamiliar ways (see e.g. Refs. \cite{Bertolami06b,Bertolami09} for discussions on 
the space-times that arise from attempts of unifying the fundamental interactions of nature and on the issue of causality).     
Indeed, developments in string theory, the most studied unification scheme, suggests that our Universe is 
actually a single element in vast multitude of about $10^{500}$ universes, a {\it multiverse} \cite{BoussoPol,Susskind}, a context  
where even the interaction between universes can be considered \cite{Bertolami07a}. 

A vast body of evidence, arising astronomy, physics, chemistry, 
geology, paleontology, biology, genetics, archeology, history, etc, support the current 
picture of Universe's history and evolution. This cosmo-vision, still under construction, is the result of the  
laborious work of several generations of intellectuals and researchers, and is one of the finest 
constructions of the human spirit. This picture allows for the understanding of the articulation of ``origins'', 
from the origin of the Universe to the emergence of life on Earth and its subsequent 
evolution till the first human social organizations. 

From studies of the Microwave Background Radiation, the surface of last 
scattering corresponding to the transition to transparency, about 370 thousands 
years after the Big Bang, one can estimate the age of the Universe as about
13.7 thousand million years (Gys). The first galaxies and stars are estimated to have formed about 
10 Gys ago. Radiative dating allows estimating the age of Earth to about 
4.5 Gys. Evidence arising from dating of stromatolites fossils suggests 
that life on Earth appeared 3.6 Gys ago. The first macroscopic fossils 
seem to have first appeared 700 million years ago. The tectonic 
processes which gave origin the Atlantic ocean took place about 100 million 
years ago and our primate ancestors walked on their inferior members about 
3 million years ago. The first human population's organizations appeared 
40 thousand years ago and the impressive first human artistic 
manifestations, the cave and rock paintings found Europe and Australia, date 
back to about 40 to 50 thousand years. 

But back to the origin of all origins, the Hot Big Bang emerges as the most successful model 
to harmonize the vast set of observations stating that the origin of the observable Universe 
took place about 13.7 thousand million years ago from a extremely hot initial state. From the epistemological point 
of view, one could argue that the strength of the model lies on the fact that it can be falsified in a great number of 
ways. Some sceptics might argue however, that the model is too flexible model as it naturally 
allows for fairly easy extensions, 
such as inflation, dark energy and dark matter, that are crucial to render the model 
consistent with the observations. If from one hand, one 
can feel uneasy with the diversity of ways the above mentioned fixes can be implemented, the fact is, on the other, 
that these fixes are suggested by a quite abundant and diverse set of observations.  

For sure, it would be quite useful if there were available contending models to the Hot Big Bang, however in the XX 
century, only the Steady State model played this role, and only for a fairly brief period. 
Thus, one can regard the Hot Big Bang model together with inflation, dark matter and dark energy are the 
paradigm and the methodological touching stone of contemporary cosmology. Epistemologically, 
the main virtue of a paradigm is that it can be falsified. For instance, the inflationary paradigm (see e.g. Ref. \cite{Olive} 
for an extensive review), has been recently challenged by the so-called ekipyrotic collisional branes proposal; however, 
given the complexities of the involved string theory (see e.g. \cite{Khoury}), few believe that the emerging picture 
is mature enough for a fair comparison. But the point here is that this alternative scenario has, in what concerns, for instance, 
gravitational waves, a distinct signature (actually, a much more modest contribution) than the one predicted by inflation. 
This might allow for a falsification criterion.    

In the context of the Big Bang, all known areas of physics (high-energy physics, nuclear physics, statistical 
mechanics, etc) are required in the process of deciphering the features of Universe's evolution. This reconstruction includes 
fairly complex and highly non-linear phenomena such as structure formation. Indeed, the Cold Dark 
Matter ``paradigm", the model based on the structure formation being triggered by non-relativistic particles 
after the decoupling with the primordial plasma, allows, through N-body simulations, to 
match the observed matter distribution on large scale and even to predict the size of the 
smallest structural seeds, actually  
Earth-mass structures \cite{Diemand}. More 
recently, simulations even allow for some understanding on the formation process of the very first stars \cite{Roseta}.

Furthermore, the Hot Big Bang model matches fairly well the observed abundance of light elements, $He^4$, $He^3$, $D$ and $Li^7$, 
that according to the model were 
synthesized a few minutes after the Big Bang, 
when temperature was about a few $MeV$, that is $10^{11}~K$. All the remaining elements had to be synthesized 
in the interiors of the stars, and the 
observation of early stars with rather few 
elements is yet another consistency check of the model.   

Of course, one could conceive that different cosmological setups might arise if instead of general 
relativity one would start from some alternative theory of gravity. However, 
the fundamental underlying principle of
general relativity, namely the connection between curvature and matter-energy as established by Einstein's 
field equations, is
consistent with all experimental evidence to considerable accuracy (see e.g. Refs. \cite{Will05,BPT06} for
reviews). Despite that, there are a good number
of reasons, theoretical and experimental, to question if general theory of relativity is the
ultimate description of gravity (see e.g. \cite{Bertolami06c} and references therein). 
These concerns are related with fundamental issues, such as 
the singularity problem, the cosmological constant problem (see e.g. \cite{Bertolami09} and references therein) 
and the underlying mechanism of inflation, 
difficulties that cannot be satisfactorily addressed in the context of general relativity. 
Therefore, it is not impossible that the Big Bang scenario may turn out 
to be inaccurate and/or fundamentally incomplete. Even though, evidence strongly suggests that this is 
not the case at all.

\section{Post-Modern Thinking: The Cosmological Dimension}
\label{sec:cosmology}

For sure, it is quite evident that technological developments have shaped and conditioned the 
evolution of humankind, most particularly from the XIX century onwards. Most likely 
this interaction will be even more intense in the future. However, it is not so clear how effectively the most fundamental 
scientific concepts, ideas and discoveries materialize themselves in day to day life. For instance, 
comparative genetic studies and 
the human genome mapping indicate that the genetic variations between the most anthropologically distant 
human groups is fairly modest. Despite that we live in a divided world and often in crisis 
due to cultural and civilizational differences. It is even less clear the impact that the cosmological discovery of 
the modest standing of humankind in the Universe, however special it might be, 
has on human culture. Another example has been pointed out by Bronowski \cite{Bronowski} and concerns  
the tragic historical circumstance of physics in 1920s coming to terms with the fact that, on the most fundamental level,  
absolute knowledge is impossible, given the limitations of the uncertainty principle in quantum mechanics, 
in a time of emergence of absolute ideological certainties. Actually, a quite dark historical period when totalitarian 
ideologies seized power over a good part of the civilized world and staged  
the most barbaric war that ever took place in the history of the humankind. A war 
that was fundamentally different from the previous ones 
as it was primarily fought on the basis of scientific and technological 
knowledge and unquestionable ideological certainties that trivialized human condition. A 
path the humankind has to avoid to repeat at all cost. 

Despite these considerations, it would not be too exaggerated to state that contemporary cosmology is an indissoluble part of 
the post-modern thinking. Indeed, science, and cosmology in particular, 
do permeate our culture, and one has already seen some of these manifestations in 
contemporary arts\footnote{For concreteness, one could mention the Italian artist Laura Pesce 
(http://www.laurapesce.it/index.html), who has 
been using scientific ideas and concepts as inspiration for her work. Conversations with this author has led her also 
to consider the ``invisible'' Universe dominated by dark energy and dark matter, as the 
object of her artistic concerns. Figure 2 depicts her view of the unifying model of 
dark energy and dark matter, the generalized Chaplygin gas \cite{Bento02}.}
science fiction, cinema\footnote{The scene of a boy in 
one of Woody Allen's films - most likely 
the director himself - who refuses to eat because the Universe is expanding is an example that strikes the memory of this author.}, 
music \cite{Caballero} and so on\footnote{One could also recall that ``Big Bang'', ``Neutrinos'', ``Gamma-Ray Bursts'', 
``Spiders from Mars'', ``Dark Matter'' and the like are the names of some rock bands; ``Supermassive Black Holes'', the 
motto of a rock song.}. 
Literature and most particularly, contemporary literature, is not indifferent to the appeal of cosmology. 
One presents some representative examples. 

\begin{figure}[]
\centering \includegraphics[width=0.5\textwidth]{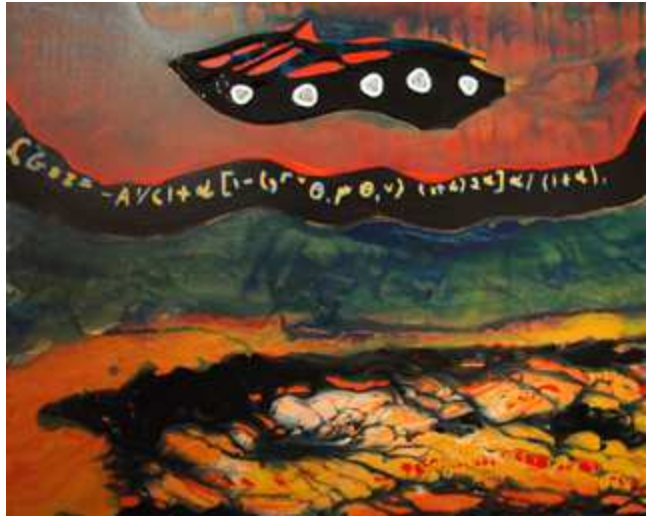}

\caption{``Dark (Energy-Matter) intertwining'' 
by Laura Pesce (2007)}

\label{fig:dark} 
\end{figure}

In ``Zuckerman unbound'', the main character of Philip Roth's (1933 --) 
novel, in a 
moving and elaborate piece of literature, uses the latest ideas about the 
origin of the Universe to comfort his dying father. Roth's character 
assumes that conveying the grandeur of the beginning of the cosmos would be a 
more effective balm rather than exposing the complexity of 
his feelings and disgust about the limitations of human existence. 

In a contribution to the Cosmology Across Human Culture meeting, which took place in 8-12 September, 2008 in Granada, 
Spain, this author discussed on the influence that cosmological thinking has had 
on the writings of authors like Italo Calvino, Ernesto S\'abato, 
Jos\'e Saramago and Fernando Pessoa \cite{Bertolami08b}.

Actually, for many authors, the Universe, with its laws and dynamics, is an active framework for the 
literary expression. Furthermore, the understanding of the machinery  of the cosmos are seen by some authors as 
guidelines for an ethics \cite{Primack}. 

Of course, the role of the allegorical in the literary exercise has always been present and the 
development of humankind has shifted the trend from the divine to the profane. Astronomy and cosmology has expanded considerably the 
boundaries of literature.  
The Moon was first visited by Cyrano de Bergerac (1619 -- 1655), in 
his ``The Other World: The Comical History of the States and Empires of the Moon", where injustice and the prominent 
anthropocentric view of man's place in creation at the time were subtly criticized. 
La Fontaine's (1621 -- 1695) fables anthropomorphized animals, plants and objects to convey moral lessons. 
In ``Gulliver's Travels" (1726) the fantastic adventures of the surgeon and captain 
Lemuel Gulliver and his contact with other ``civilizations" in remote parts of the world, was the way found by Swift (1667 -- 1745)) 
to examine human shortcomings and to 
criticize the society of his time. Inadequacy could not be more expressively depicted than through 
the ``Verwandlung" of Gregor Sansa into an insect. Indeed, Kafka's (1883 --1924) metamorphosis of Gregor Sansa 
has inspired many authors who discovered that there should be no limit to their imagination and the door of 
opportunity was then wide open to works of art such as Philip Roth's ``The Breast" (1972), the work of the 
widely acclaimed contemporary Japanese author Haruki Murakami (1949 --), etc. Literary surrealism, 
whose roots can be possibly traced back to ``Hieroglyphic tales" (1785) of Horace Walpole (1717 -- 1797), 
has put all the emphasis on the subjective and on the unconscious world,  but its scope 
has always been the analysis of the deepest human in a Universe ruled by well defined and stable laws. Understanding 
the ``origin and formation of reality", was the declared purpose of the novel ``Kosmos" (1965) 
by Witold Gombrowicz (1904 --1969). A novel crowed with disconnect and unlikely events, 
even though all of them possible in the realm of the physical world.    
Most likely, this is the common pattern of all fantastic literature, which deliberately introduces  
supernatural elements in a world that evolves according to the physical and sociological laws. 
Most likely, this is the most effective way to 
explore and develop contradictory conceptual elements, the quintessential function of a genuine work 
of art\footnote{One refrains from bringing examples from science fiction, 
as in this genre the connection with science is rooted from the very start.}.

\section{A Cosmic Responsibility}
\label{sec:conclusions}

In ancient cultures, historical development was seen as 
an extension on the human sphere of a cosmogony which took place in the 
natural world; given that the latter was due to a divine intervention, it 
leads to an umbilical relationship between religion and ethical values. 
Thus, cosmology and religion were once genetically connected. 
This is fairly clear in the context of the great religions, a connection that can be found in many cultures.  

Therefore, an interesting question is whether one could envisage a 
``cosmic ethics" without relating it to a religious view of the world. That is, could one conceive taking 
cosmology as the cornerstone for an ethics based on values such as {\it universal harmony} and {\it responsibility}?. 
A strictly scientific answer can only be negative, given that  
scientific advances and discoveries were achieved independently from humanistic and anthropocentric concerns. 
The scientific facts that describe and allow for the understanding local dynamics is a 
limit of a general set of laws that govern the Universe, and hence, it is 
improper to ask for the ethical implications that research on the infinitely large and 
on the infinitely small might have. Furthermore, cosmology provides an eloquent perspective of 
the modest standing of humankind in the cosmos. Nevertheless, cosmology renders a 
view of how unique, and this is a somewhat anthropocentric interpretation, are the conditions required to 
shelter life, and particularly sentient and reflective life. 
Even if life is a wide spread phenomenon in the Universe, a firm believe of this author, humankind is most likely, 
quite unique within the family of 
self-conscious species that exist throughout the Universe. In either case, human species has the collective  
responsibility of keeping the balance of the world and to ensure its continuity. 

Similar thoughts were raised by Hannah Arendt, back in 1963, when reflecting upon the ethical 
implications of the conquest of space \cite{Arendt}. How is humankind enlightened by the possibility 
of being able to go beyond its ancestral home was then the question. On a broader sense, 
cosmology has acquainted humankind with events at the very edge of the observable Universe. It has expanded our 
scale of space and time beyond anything that could be envisaged from the observation of 
the nearby world. But, the question is if cosmology on its own is able to provide the necessary 
driving force to be the basis for an ethical system of values. Values based on the likelihood 
that one might have at last, through the Hot Big Bang model or actually any other theory, 
a considerable grasp of the inner workings of Universe's dynamics \cite{Primack}. This author is sceptical whether this purpose can 
be achieved on purely scientific grounds and through a scrupulous, yet inevitably judicious, 
exhibition of the facts. Indeed, if there is any chance of reaching an universal 
ethical code, this has to be based on the history of humankind itself. 
It should be based on the lessons drawn from the heroic development 
of humankind against oppression, irrational forces and dogma. A centenary, and still unfinished 
struggle, where reason and science, cosmology included,  
play a crucial role. 

Even though the future might not be more than ``a structure of hopes and expectations, whose residence is in the 
mind and has no reality" \cite{Coetzee}, one feels that future trends are taking shape and that these 
are continuously emerging from current tendencies, developments, crises and catastrophes. 
It is a believe shared by 
many, that what future might have on stock is closely related with the above mentioned struggle. 
One has reached a time when the inability 
of the institutions, national and international, to tackle the urgent needs and problems of humankind 
does give room to the development of forces that question the very foundations of 
what made possible the noblest achievements of the human   
civilization. Currently one witnesses a dangerous clash between obscurantism and humanist values. By these one means the set of 
values that allow for a rational and scientific view of reality in societies based on equality civil rights and democracy. For sure, 
values that act upon reality through imperfect constructions and foundations, but that are, nevertheless, 
the most effective way to ensure the well being of the human community. Values that 
allow for regarding societies with a critical eye and to fix their mistakes. Of course, it would be most absurd to deny 
that social injustice exist also in democratic societies. It would be intellectually dishonest not to see that science has 
greatly multiplied the power to alter the environment and that this power may turn out to be a major threat 
for humankind's future. Despite that, abandoning altogether humanistic values would bring even more serious risks. 
Without science and democracy, the world would be even more exposed to unhappiness and hopelessness. 
Unfortunately, the fate of these values, and in a way the very future of humankind, is not yet free 
from danger.  At least three major causes for concern can be identified: 

\ni
1. The inability to properly tackle social injustice and poverty, which besides the problems they pose, 
lead in some parts of the world to the fragmentation 
of the state and the breakdown of all cultural and civilizational values. This collapse pushes civilization back to the rule 
of the arbitrariness and dogma. It is unquestionable that the 
continuation of these conditions is a threat to the stability of the whole world.    

\ni
2. The recent arousal of religious fundamentalism, a phenomenon that is now emerging through various guises 
in many parts of the world, including in the most developed countries, does represent a serious challenge to the humanist 
set of values discussed above. One does not need to go as far as to state that faith is the enemy of humanism, even though 
it can be argued that it menaces science (see e.g. \cite{MacKenzie}). But it is undeniable that any form of religious or political 
fundamentalism 
is incompatible with the intellectual tolerance that is essential for the development of culture, science, human equality and 
solidarity. 

\ni
3. One has been witnessing, most particularly in the most developed parts of the world, a 
sharp and worrisome fall of the cultural and educational standards. This decline breeds a culture of intolerance, 
of misunderstanding and 
mistrust of cultural and intellectual achievements. It nurtures a culture based on blind and selfish materialism and 
consumerism. Needless to say that this social trend puts at risk social stability and  
imperils the balance of the environment and ecosystems.

It would be a mistake to assume that the above discussed humanistic values are a one way road towards a 
globalized and culturally homogeneous world. If so, that would impoverish human existence and 
make the world intellectually uninteresting and eventually sterile. 
Humanism is the result of a historical evolution which involved the whole humankind. It is 
perfectly consistent with a multi-cultural and multi-centered world. The diversity of the human experience is at the 
very foundation of a system of values that puts the integrity of the human life, and actually of all life on our planet, above 
everything else. As emphatically expressed by George Steiner in ``After Babel''  \cite{Steiner1}, the true tragedy of   
Babel is not the scattering of languages, but the reduction of human speech to a handful 
of planetary, ``multinational" tongues. A reduction impelled 
by market forces and information technology, and that dangerously threaten the survival of some 
languages and of the human culture on a broad sense. ``When a language dies, it is not only the vital lineage of remembrance -- 
past tenses or their equivalence -- it is not only a landscape, realistic or mythical, calendar, 
which are blotted out: it is configuration of a conceivable future"  \cite{Steiner2}.
Humanism can only thrive and fully blossom in a multi-cultural world.

In what concerns this point, it is relevant to remember that a multi-cultural society existed till the XIII century in Europe. 
It was centered in Cordoba and was the result of the historical development of 700 years of Muslim ruling of the 
Iberian Peninsula. In Cordoba, individuals of different faiths 
excelled in science, philosophy, theology, agriculture, art, and architecture. Their lasting contributions 
created an advanced and thriving center of innovation for both material culture and sciences. In Cordoba, 
leading intellectuals and scholars worked in medicine, urbanism, astronomy and philosophy. They 
translated and scholarly commented on the mathematical and philosophical achievements of the 
Hellenic culture. It is through these texts and studies, most often in Arabic and sometimes also in Hebrew, 
that the richness of the pre-Socratic thinking, and the work of Plato and Aristotle have survived and reached modern times.
In Cordoba, Muslims, Catholics and Jews lived side by side and built a quite sophisticate society 
which gave rise to important contributions that shaped the prevailing philosophical and theological 
thinking for several centuries. Two names stand out, given the depth and the prevalence of their contributions: 
Averroes and Maimonides. 

From a family of Muslim Andalusian scholars, Ibn Rushd, Averroes as he is known in the Western world, 
was born in Cordoba in 1126 and died in Marrakech in 1198. His extensive comments on Aristotle's work (but the ``Politics") 
and on Plato's ``Republic" are a fundamental starting point for the understanding of those Greek authors. 
In his ``Incoherence of the Incoherence", his most original philosophical contribution, Averroes argued, 
in opposition to previous Muslim scholars, that Aristotle was not self-contradictory and that his ideas were 
not against the teachings of Islam. He also believed that there was no conflict between philosophy and religion, 
which he regarded as complementary paths to reach the truth.  
Averroes contributions to Astronomy were also relevant. He rejected the eccentric deferents introduced by Ptolemy and 
defended a concentric planetary system and Universe. He has also made the first descriptions of sunspots and of the 
reflection of sun's light on the surface of the moon. For Averroes, the Universe was eternal. 

In medicine, Averroes discussed dissection and autopsy, although he has never performed any 
of them himself. He supported that their practice as a way to ``strengthen the faith" 
as it allowed to observe ``the remarkable handicraft of God in his creation". He also diagnosed the 
Parkinson's condition and suggested that the photo-reception was the main feature of the retina, 
which according to him was the central organ of sight. 

Moshe ben Maimon, but better known as Maimonides, was also born in Cordoba, in 1135. 
He was a rabbi, physician and philosopher who also lived briefly in the holy land, in Morocco 
and passed away in Fostat, Egypt in 1204. 
He is associated with the end of the golden age of Jewish orthodox culture and his rationalism and 
strong opposition to mysticism exerted substantial influence in non-Jewish scholastic 
philosophers like Albert the Great, Thomas Aquinas and Duns Scotus. In Egypt, he was the physician of 
Grand Vizier Alfadbil and Sultan Saladin. He also treated Richard Lionheart while on the Crusades. 
Maimonides wrote extensively on Jewish scholarship, rabbinic law, philosophy and several medical texts 
(Treatise on Poisons and their Antidotes, Treatise on Regimen and Health, 
Treatise on Causes and Symptoms, Treatise on Cohabitation, Laws of Human Temperament and Treatise on Asthma), 
most of them in Arabic. His most well known work in Hebrew, the Mishnet Torah (Second Torah) 
comprises a code of Jewish law and the fundamental 13 principles of faith (Existence of God, Unity of God, 
God's eternity, spirituality and non-corporeal nature, etc), and constitutes the foundations of orthodox Judaism. 
In his extraordinary ``Guide for the Perplexed'', Maimonides confronts the spatial 
infinity and the eternal cosmos of Aristotle with the Torah's 
cosmogony which presumes a genesis for the start of the world, 
a contradiction which was never properly acknowledged by Christian scholastics.

Just 160 km from Cordoba, stands another multi-cultural town, Granada.  
Originally, Jews and Muslims emigrated to Granada, from the nearby town of Elvira, 
by the VIII century. At the time, most of the Iberian Jews inhabited the area where 
Muslims immigrants began developing the city at the base of the Sierra Nevada Mountains. In mid XIII century king 
Fernando III conquered many cities, including Muslim-ruled Seville and Cordoba. To prevent the 
Christian king imminent invasion, Granada ruler Muhammad Ibn Ahmar made a treaty which consisted in the payment of 
an annual tribute and in providing assistance to Fernando III military campaigns. Ibn Ahmar and his descendants 
ruled the kingdom of Granada for more than two centuries. 
Throughout their reign, Muslim and Jewish refugees arising from cities conquered by Christians flocked the Granada. 
The city was the last Muslim kingdom on the peninsula and eventually, on January 2nd, 1492, 
Isabella and Ferdinand, the Catholic Kings, forced its last Muslim ruler, Boabdil, to surrender. 
  
Alhambra stands as a magnificent remainder of the most glorious days of Muslim Europe and the adjoint 
palace of Carlos V is an example on how different cultures can live side by side despite their differences. 
Alhambra is among the most remarkable monuments of the Muslin architecture in the Western world and 
the multi-cultural and historical conditions which gave rise this most magnificent 
construction can be regarded as a particularly appealing model for the future of humankind.

\section{Our Cosmic Future}

``Voyages dans les futur'' by the astrophysicist Nikos Prantzos \cite{Prantzos} is a quite stimulating reading about scenarios, 
based on science and on science fiction literature, for the future of our civilization. From plans to get 
back to Moon to the research on new propulsion methods to explore the solar system and beyond, the reader is 
wrapped by the irresistible 
call of exploration that impels our species towards space. If from one hand, cosmology provides the measure that allows one 
to understand the modesty of any such undertaking, on the other, it urges one to deepen the  
quest for knowledge. An endless and humble quest, but by all means, an inevitable one. 

Even though space exploration is nowadays an essentially scientific goal, in the future it 
might turn out to be ecological and economical imperatives.   
The growth of the human population and the finiteness of resources of our planet might require the search for new habitats 
and raw materials. However, in order to achieve that scientific and technological 
knowledge must be considerably advanced, most particularly in what concerns methods of propulsion in space. Special relativity 
does provide an objective way to gauge technological achievement through 
comparison with the speed of light, $c=300,000~km/s$, the ultimate speed limit. In what concerns elementary particles, 
high-energy physics colliders allow accelerating particles up to $99;9999\%$ of the speed of light; 
however, in what refers to macroscopic objects, and spacecraft in particular, science is still fairly rudimentary as the 
greatest velocity ever achieved is Earth's scape velocity, that is $11.2~km/s$ or $3.6 \times 10^{-5} c$. 
Indeed, space exploration necessarily involves the release of considerable amounts of energy  kept in fairly compact devices. 
Using special relativity once again, the well known equation $E=mc^2$ provides the metric to measure 
the process of mass conversion into energy. The rocketry of our days, based on chemical propulsion, involves the release of 
chemical energy that corresponds to a conversion of $10^{-10}$ of the initial mass, yielding ejection velocities of 
propellants up 
to about $10~km/s$. Nuclear energy methods are millions of times more efficient, and in fission reactions, a conversion 
factor of about $7 \times 10^{-4}$ can be achieved. In the process of fusion of the hydrogen isotopes, a conversion factor of 
about $5 \times 10^{-3}$ can be reached. At least theoretically, these conversion factors might allow reaching ejection 
velocities of about $0.01c$ to $0.1c$. 

It should be kept in mind that the propulsion process is based on Newton's third law, according to which {\it action and reaction 
are equal and opposite to each other}, so that linear momentum carried by the ejected propellant is transmitted to the rocket. 
The relationship between the final velocity of the rocket, $v$ and the ejection velocity of the propellant, $v_p$ is given by the 
Tsiolkovsky or rocket equation, $v=v_p \ln(M/m)$, where $M$ is the initial mass of the rocket and $m$ its mass when the velocity is 
$v$. Konstantin Eduardovitch Tsiolkovsky was the Russian visionary and mathematics teacher who in 1897 deduced this equation. 
In his 1903 book 
``Cosmic Space Exploration with Reaction Engines", he discussed for the first time propulsion based 
on a mixture of liquid hydrogen and oxygen, multistage rockets, space suits and attitude control through the use of gyroscopes 
among many other daring and revolutionary ideas. His statement: ``Our planet is the cradle of intelligence, but one cannot 
live forever in a cradle" summarizes the believe, shared by many, that space exploration is an inevitable implication of the 
human development.    
Inspired by Tsiolkovsky, scientists, engineers, science fiction authors have come up with exciting ideas for new methods of 
propulsion which involve for instance, matter-antimatter annihilation, space sailing using solar radiation, navigation impulse 
produced by intense laser or microwave beams, spacecraft that gather interstellar hydrogen for its nuclear fusion engine, etc.       

One the most imaginative ways ever conceived 
to reach space is the one that involves the control of gravity itself. Suggested by H.G. Wells in 
1901 in his book ``The First Men in The Moon", a rather unlikely crew reached its destination using a spacecraft made out of a 
material endowed with anti-gravity properties, the cavorite. Since then the possibility of switching off gravity has been 
discussed in the science fiction, Internet groups and more rarely in the scientific literature. Sparked by the interest of Boeing and 
of NASA, study programmes where setup to examine this and other innovative ideas for propulsion, NASA's Breakthrough Propulsion 
Project (1996 -- 2002) and the British Greenglow Project (1997 -- 2002) being the most well known. 
The European Space Agency (ESA) has sponsored a study on the subject of gravity control which involved this author and 
Martin Tajmar \cite{Bertolami2002a,Bertolami2002b}, and its conclusions were, as expected, negative. Indeed, it 
has been throughly argued and discussed that gravity cannot be controlled, and even if this were feasible, only under quite 
particular conditions it would be more effective than conventional means of propulsion \cite{Bertolami2002a,Tajmar2005}. Even though 
these findings were obtained in the context of Newtonian mechanics, subsequent work has shown that conclusions would 
be essentially similar even after general relativistic considerations \cite{Bertolami2007}. Furthermore, in the ESA's study 
other means of cosmic traveling such as wormholes, warp-drives, etc were also critically assessed and general arguments based on the 
positiveness of the energy were used to argue that these space-time shortcuts are most likely unstable and physically 
unfeasible. 

Of course, besides the limitation of the laws of physics, one has also to consider the vastness of the cosmos, 
which prevents any foreseeable exploration undertaking much beyond our most immediate stellar neighbourhood. 
However, this might not be the final word on the human exploration of the cosmos as one could instead envisage a 
robotic exploration based 
a self-replicating robot spacecraft, a ``von Neumann probe'' \cite{Boyce}. These probes would consist of a propulsion system and 
a universal von Neumann intelligent replicator, to be launched toward a neighbouring stellar system. Upon arrival it 
would seek out raw materials, from local sources and use them to make several copies of its hardware. 
The copies would then be launched to the following set of neighbouring stars. The replication of the process and 
the continuous increase of probes would allow exploring ever more remote regions of the Galaxy. Actually, it has been argued by 
Tipler \cite{Tipler} that the von Neumann probe approach is so logical and economical that it would be adopted by any 
advanced civilization. He estimated that if probes could reach velocities of up to about $0.1~c$, a complete galactic colonization 
could be achieved in about 200 million years, less than $5 \%$ of the age of the Galaxy. From the fact that no such devices have been 
detected, he draws the conclusion that that humans are the only intelligent species among the Galaxy's several hundred billion stars. 
The validity of the argument has however, been questioned by Sagan and Newman \cite{Sagan}, on the basis that if the growth of the 
number of probes were exponential, a single self-replicating probe could be expected to convert the entire mass 
of the Galaxy into copies of itself within 2 million years! Therefore, concluded Sagan and Newman, any species intelligent 
enough to build such probes would realize the danger of such an ``infectious'' project. Of course, these arguments refer to 
civilizations that are much more advanced technologically, and may not concern humankind, at least in the foreseeable future. However, 
through for instance the exploration of Mars, despite the model scale of this achievement, one could draw conclusions about 
the advantages and the relative 
simplicity of employing robotic based strategies in the space exploration. It would not be an unthinkable stretching of imagination to 
conceive going beyond the solar system using a robotic based approach. The development of self-replicating von Neumann machines would 
be then the next technological hurdle to master. 

At this point it is rather natural to speculate about the possibility of detecting evidence for life. For this author, arguments 
based on the absence of evidence are not quite satisfactory as they do not necessarily imply any evidence of absence. 
A somewhat more pleasing argument is the one that arises from the universality of the laws of 
physics, and hence of the laws of chemistry. From this universality once can conclude that from a set of favourable conditions, 
such as the ones found on Earth, the emergence 
of life must be a rather common occurrence given the vastness of the Universe. How common is still an open question. 
However, one might not have to wait too long to get a clearer idea of the answer  
once it is understood how general is the occurrence of planets around the stars, to what degree the so-called habitable 
zone can be extended to the satellites of large gaseous planets, what are the most reliable markers to detect the presence of life, 
etc. An exciting prospect, but till then, human civilization has to face 
the immediate challenge of coping with the menace it posed itself by the misuse of its own planet. Alas wisdom prevails.

\vspace{0.3cm}

{\bf Acknowledgments~~}

\noindent
The author is indebted to Jorge P\'aramos and Carlos Zarro for their constructive comments and suggestions.



\bibliographystyle{unstr}

\end{document}